\documentclass[prd,amsmath,10pt,superscriptaddress,floatfix,nofootinbib]{revtex4-2}
\usepackage{epsfig,amssymb,amsfonts,amsmath,mathtools,bm,color,xcolor,graphicx,braket,adjustbox,esint,upgreek}
\usepackage{multirow}
\usepackage[utf8]{inputenc}
\usepackage{setspace}
\usepackage{dcolumn,kantlipsum}
\allowdisplaybreaks

\usepackage[
colorlinks=true,
linkcolor=blue,
breaklinks=true,
urlcolor=blue,
citecolor=blue]{hyperref}

\usepackage{orcidlink}
\usepackage{mathrsfs} 

\usepackage{bm,bbm}
\usepackage{slashed}

\newcommand{\be}{\begin{equation}}
\newcommand{\ee}{\end{equation}}
\newcommand{\bea}{\begin{eqnarray}}
\newcommand{\eea}{\end{eqnarray}}
\newcommand{\beas}{\begin{eqnarray*}}
\newcommand{\eeas}{\end{eqnarray*}}

\renewcommand{\arraystretch}{1.2}
\def\vec#1{\boldsymbol{#1}}
\newcommand{\nn}{\nonumber}

\renewcommand{\L}{\mathscr{L}}

\newcommand{\MeV}{\,\text{MeV}}
\newcommand{\GeV}{\,\text{GeV}}
\renewcommand{\vec}[1]{\mathbf{#1}}
\newcommand{\diff }{{\text{d}}}

\newcommand{\ustb}{\affiliation{School of Mathematics and Physics, University
of Science and Technology Beijing, Beijing 100083, China}}

\graphicspath{{FiguresV2/}}

\begin{document}

\title{ Nature of the $X(2370)$: A light-quark perspective }

\author{Yun-Hua Chen\orcidlink{0000-0001-8366-2170}}\email{yhchen@ustb.edu.cn}
\ustb

\begin{abstract}

\medskip

The BESIII Collaboration recently claims that the $X(2370)$ is a pseudoscalar glueball candidate, and one of the most important properties of glueball is that it is a flavor singlet.
In this paper, we try to reveal the structure of the $X(2370)$ from the light-quark perspective.
We study the $K_S^0 K_S^0 \eta^{(\prime)}$, $K_S^0 K_S^0$ and $K_S^0 \eta^{(\prime)}$ invariant mass distributions and the angular distribution of $K_S^0$ in the $K_S^0 K_S^0$ rest system of $J/\psi \to \gamma K_S^0 K_S^0 \eta^{(\prime)}$ processes and the decay-width ratio of $\Gamma(X(2370) \to K\bar{K}\eta^\prime)/\Gamma(X(2370) \to \pi\pi\eta^\prime)$. In
particular, we consider the effects of different light-quark SU(3) eigenstates inside the $X(2370)$. The strong $\pi\pi$-$K\bar{K}$ coupled channel final-state interactions in the $S$- and $D$-waves
are taken into account model independently by using dispersion theory. We find that the light-quark SU(3) octet state of the $X(2370)$ plays a significant role in these transitions.
Furthermore, we find that if only consider the flavor singlet of $X(2370)$, the decay widths ratios ${\Gamma(X(2370) \to K\bar{K}\eta^\prime)}/{\Gamma(X(2370) \to \pi\pi\eta^\prime)}$ and ${\Gamma(X(2370) \to \pi\pi\eta)}/{\Gamma(X(2370) \to \pi\pi\eta^\prime)}$ can not be described simultaneously. Our study indicates that the $X(2370)$ contains a
large light-quark component and thus cannot be a pure glueball. Also, we finds that in the $X(2370)$ decays into three pseudoscalars, $X(2370) \to PPP$, the energy-dependent couplings should not be neglected.

\end{abstract}

\maketitle
\medskip
\medskip
\medskip

 \begin{onecolumngrid}

\section{Introduction}\label{intro}

The nature of the $X(2370)$ structure is still in heated debate since its discovery
by the BESIII Collaboration in 2011 in the process $J/\psi\to\gamma\pi^+\pi^-\eta'$~\cite{BESIII:2010gmv}. It was subsequently confirmed by an analysis of both charged and neutral $J/\psi\to\gamma K\bar K\eta'$ modes~\cite{BESIII:2019wkp}. A partial-wave analysis of the $J/\psi\to\gamma K_S^0K_S^0\eta'$ process determined its spin parity to be $J^{PC}=0^{-+}$~\cite{BESIII:2023wfi}. 
The self-interaction of gluons in Quantum Chromodynamics (QCD) permits the existence of glueballs~\cite{Amsler:2004ps,Crede:2008vw}, and lattice calculations provide predictions for the ground state of a pseudoscalar glueball with a mass around $2.3 - 2.6~\mathrm{GeV}$/${c}^{2}$~\cite{Bali:1993fb,Morningstar:1999rf,Chen:2005mg,Gregory:2012hu,Gui:2019dtm}. Recently, BESIII reported that there is no significant $X(2370)\to K^*(892)\bar K$ contribution in the $J/\psi\to\gamma K_S^0K_S^0\pi^0$ process, and interpreted this suppression as evidence for a flavor-singlet assignment of $X(2370)$~\cite{BESIII:2026mvn}. BESIII considered that a dominant pseudoscalar-glueball component is essential for the explanation of the properties of the the $X(2370)$, including its 
mass, spin-parity, and production and decay properties~\cite{BESIII:2026mvn}.
While on the theoretical side, competing models have been proposed to interpret the
$X(2370)$ as a conventional pseudoscalar excitation~\cite{Yu:2011ta,Peng:2026yol}, a glueball~\cite{Tan:2026uue,Tan:2026crv,Wang:2026iyq,Giacosa:2026pqh,Buisseret:2026acu,Li:2024fko}, a glueball-charmonium mixing state~\cite{Chen:2026lki,Zhu:2026zju}, a hadronic molecule of $\Sigma\bar\Sigma$~\cite{Ben:2026afy} or $\phi h_1(1380)$~\cite{Khemchandani:2026syv}, a multiquark configuration~\cite{Deng:2012wi,Wang:2025nme}, a hybrid~\cite{Lu:2026foy}, coupled-channel dynamics~\cite{Dong:2026smv}, and so on~\cite{Sun:2021kka}.

In this work, we will study the possible light-quark components of the $X(2370)$ to help reveal its internal structure. We will focus on the $J/\psi \to \gamma K_S^0 K_S^0 \eta^{(\prime)}$ processes, of which BESIII has measured the $K_S^0 K_S^0$ and $K_S^0 \eta^{(\prime)}$ invariant mass distributions and the angular distribution of $K_S^0$ in the $K_S^0 K_S^0$ rest system~\cite{BESIII:2015xco,BESIII:2023wfi} and thus allows us to extract the information of the light-quark SU(3) flavor-singlet and flavor-octet components of the parent state of $K_S^0 K_S^0 \eta^{(\prime)}$. 
Note that the invariant mass distributions and the angular distributions given in BESIII papers~\cite{BESIII:2015xco,BESIII:2023wfi} do not distinguish the contributions of different parent states, and therefore we take account of the effects of the candidate parent states $X(1835)$, $X(2370)$, and $\eta_c$, and also fit the $K_S^0 K_S^0 \eta^{(\prime)}$ parent-mass spectra.  
The decay-width ratio of $\Gamma(X(2370) \to K\bar{K}\eta^\prime)/\Gamma(X(2370) \to \pi\pi\eta^\prime)=\Gamma(J/\psi \to \gamma X(2370) \to \gamma K\bar{K}\eta^\prime)/\Gamma(J/\psi \to \gamma X(2370) \to \gamma \pi\pi\eta^\prime)$~\cite{BESIII:2026mvn} is relevant to the strange-quark component, and will also be taken into account. If the $X(2370)$ contains no light quarks (as in the glueball or the hybrid state scenarios), the light-quark source provided by the $X(2370)$ has to be in the form of an SU(3) singlet state. Thus the determination of the contributions from different SU(3) eigenstate components is useful for clarifying the internal structure of the $X(2370)$, especially when a nonzero SU(3) octet component is found to be essential for reproducing the experimental data.

The conservation of parity and $C$-parity require the dikaon(dipion) system in $J/\psi \to \gamma  X(2370)/X(1835)/\eta_c \to \gamma  K\bar{K}\eta^{(\prime)} (\gamma  \pi\pi\eta^{(\prime)})$
to be in even partial waves. There are strong coupled-channel final-state interactions (FSIs) in the dikaon(dipion) system, which include the scalar resonances $f_0(500)$ and $f_0(980)$
in the $S$-wave, and include the $f_2(1270)$ resonance and $f_2^\prime(1525)$ resonance in the $D$-wave. The FSIs can be taken into account model-independently using dispersion theory, which satisfying unitarity and analyticity. At low energies, the amplitudes should agree with the leading chiral results. Therefore, we construct the chiral contact Lagrangians for the $X(2370)/X(1835)/\eta_c  K\bar{K}\eta^{(\prime)} (\pi\pi\eta^{(\prime)})$ couplings, and in which the parameters are then fixed from fitting to the BESIII data.

This paper is organized as follows. In Sec.~\ref{theor}, we describe
the theoretical framework and present the calculation of the
amplitudes as well as the dispersive treatment of the FSI. In Sec.~\ref{results}, we present the fit results
and discuss the light-quark components of the $X(2370)$ and its
structure. A summary is given in Sec.~\ref{summary}.

\medskip

\section{Theoretical framework}\label{theor}
\subsection{Lagrangians}

In general, the light-quark sources provided by the $X(2370)/X(1835)$ in the $X(2370)/X(1835) \to PPP$ transitions may come from two ways, one is from the possible light-quark components contained in the $X(2370)/X(1835)$ (e.g. in the $\Sigma\bar\Sigma/\Lambda\bar{\Lambda}$ molecule or the fourquark scenarios), and the other way is that the light quarks are excited by the $X(2370)/X(1835)$ from vacuum (e.g. in glueball or hybrid scenarios). Here we do not distinguish these two types of possible light-quark sources, but consider them in an unified scheme, since what is matter here is the relative strengths between the light-flavor SU(3) singlet part and SU(3) octet part acting in these transitions. Considering the light quarks sources provided by the $X(2370)/X(1835)$ in the $X(2370)/X(1835) \to PPP$ transitions, the $X(2370)/X(1835)$ can be decomposed into SU(3) singlet and octet components of light quarks,
\be
\label{eq.YComponents} |X(2370)/X(1835)\rangle=a^{X(2370)/X(1835)}|V_1\rangle+b^{X(2370)/X(1835)}|V_8\rangle\,,
\ee
where $|V_1\rangle \equiv 
\frac{1}{\sqrt{3}}(u\bar{u}+d\bar{d}+s\bar{s})$ and $|V_8\rangle \equiv
\frac{1}{\sqrt{6}}(u\bar{u}+d\bar{d}-2
s\bar{s})$.
Expressed in terms of a $3\times3$ matrix in the SU(3) flavor space, the $X(2370)/X(1835)$ is written as
\begin{equation}
 \frac{a^{X(2370)/X(1835)}}{\sqrt{3}} V_1 \cdot \mathbbm{1}+\frac{b^{X(2370)/X(1835)}}{\sqrt{6}} V_8\cdot \text{diag} \left( 1,  1, - 2\right)    .
\end{equation}

The effective Lagrangian for the $X(2370)/X(1835)PPP$ contact couplings, at leading order in the chiral expansion,
reads~\cite{Cirigliano:2006hb,Chen:2019mgp,Mannel:1995jt,Chen:2015jgl,Chen:2016mjn,Chen:2019gty}
\begin{eqnarray}\label{LagrangianXPPP}
\L_{XPPP} =&& g_1^{X}  V_{1}^{X} \langle h^{\mu\nu} u_\mu
u_\nu\rangle   +i h_1^{X} V_{1}^{X} \langle \chi_+\rangle \langle \chi_-\rangle  +g_{8a}^{X} \langle  V_{8}^{X} \{h^{\mu\nu}, u_\mu
u_\nu\} \rangle    +g_{8b}^{X} \langle  V_{8}^{X} u_\mu h^{\mu\nu} u_\nu \rangle \nn \\ && +i h_8^{X}  \langle V_{8}^{X}\chi_+\rangle \langle \chi_-\rangle
+\mathrm{H.c.}\,,
\end{eqnarray}
in which ``$X$'' can denote $X(2370)/X(1835)$, and $\langle\ldots\rangle$ represents the trace in the SU(3) flavor space.

In the large $N_C$ limit, the $U_A(1)$ anomaly in QCD is suppressed and therefore the singlet
$\eta_0$ meson becomes the ninth pseudo-Goldstone boson and can be systematically
incorporated into the $U(3)$ chiral Lagrangian~\cite{Coleman:1980mx,Herrera-Siklody:1996tqr,Kaiser:2000gs}.
We use the exponential parametrization for $U(3)_L \times U(3)_R/ U(3)_V$ coset coordinates
\begin{eqnarray}
U &=& u^2 = e^{i\frac{ \sqrt2\Phi}{ F}} \,,
\end{eqnarray}
where the pseudo-Goldstone boson octet plus singlet are given by
\begin{equation}
\Phi=
 \left( {\begin{array}{*{3}c}
   {\frac{1}{\sqrt{2}}\pi ^0 +\frac{1}{\sqrt{6}}\eta _8+\frac{1}{\sqrt{3}}\eta_0 } & {\pi^+ } & {K^+ }  \\
   {\pi^- } & {-\frac{1}{\sqrt{2}}\pi ^0 +\frac{1}{\sqrt{6}}\eta _8+\frac{1}{\sqrt{3}}\eta_0} & {K^0 }  \\
   { K^-} & {\overline{K}^0 } & {-\frac{2}{\sqrt{6}}\eta_8+\frac{1}{\sqrt{3}}\eta_0 }  \\
\end{array}} \right) \,.
\end{equation}
Here $F$ corresponds to the pion decay constant $F_\pi=92.1\MeV$ in the chiral limit.
The basic building blocks involving the pseudo-Goldstone bosons and external source fields read
\begin{eqnarray}\label{defbb}
&& u_\mu = i \left( u^\dagger\partial_\mu u\, -\, u \partial_\mu
u^\dagger\right) \,,   \qquad \qquad  h_{\mu\nu}=\partial_\mu u_\nu +\partial_\nu u_\mu\, ,  \nn\\
&& \chi_\pm  = u^\dagger   \chi u^\dagger \pm  u  \chi^\dagger  u \,,
 \end{eqnarray}
where $\chi=2 B_0 (s + i p)$ incorporates the pseudoscalar ($p$) and scalar ($s$) external sources.

The physical states $\eta$ and $\eta^\prime$ are from the mixing between $\eta_8$ and $\eta_0$.
Following a general discussion in $U(3)$ $\chi$PT, the $\eta$-$\eta^{\prime}$ mixing should be expressed in the
two-mixing-angle framework, instead of the conventional one-mixing-angle description~\cite{Leutwyler:1997yr}.
In the octet and singlet basis, the $\eta$-$\eta^{\prime}$ mixing is parameterized as~\cite{Leutwyler:1997yr}
\begin{eqnarray} \label{twoanglesmixing81}
 \left(
 \begin{array}{c}
 \eta   \\
 \eta^\prime \\
 \end{array}
 \right) = \frac{1}{F}\left(
                                        \begin{array}{cc}
                                          F_8\, \cos{\theta_8}  & -F_0 \,\sin{\theta_0}  \\
                                           F_8\,\sin{\theta_8} & F_0 \,\cos{\theta_0} \\
                                        \end{array}
    \right)
      \left(
       \begin{array}{c}
       \eta_8   \\
       \eta_0  \\
       \end{array}
        \right)\,,
\end{eqnarray}
where $F_8$ and $F_0$ represent the weak decay constants of
the axial octet and singlet currents, respectively. By taking
$F_8=F_0=F$ and $\theta_0=\theta_8$ in Eq.(\ref{twoanglesmixing81}),
the conventional one-mixing-angle description is recovered. For the values of $F_8$, $F_0$, $\theta_8$, and $\theta_0$, we will use the results in~\cite{Chen:2014yta}.

We also need the effective Lagrangian for the $\eta_c PPP$ contact coupling, which only contains two terms since $\eta_c$ is a flavor-singlet state
\begin{eqnarray}\label{LagrangianetacPPP}
\L_{\eta_c PPP} =&& g_1^{\eta_c}  \eta_c \langle h^{\mu\nu} u_\mu
u_\nu\rangle +i h_1^{\eta_c} \eta_c \langle \chi_+\rangle \langle \chi_-\rangle +\mathrm{H.c.}\,.
\end{eqnarray}

The gauge-invariant Lagrangian for the $J/\psi \to \gamma X(2370)/X(1835)/\eta_c$ coupling is given by
\begin{eqnarray}\label{LagrangianJpsigammaXoretac}
\L_{\gamma} =\epsilon^{\mu\nu\alpha\beta} \partial_\mu \psi_\nu F_{\alpha\beta}\big(g_\gamma^{X(1835)}X(1835)+ g_\gamma^{X(2370)}X(2370)+g_\gamma^{\eta_c}\eta_c\big)+\mathrm{H.c.}\,.
\end{eqnarray}

\subsection{Amplitudes of \boldmath{$ X(2370)/X(1835)/\eta_c \to \eta[\eta^\prime] \pi\pi(K\bar{K}) $} processes}

The decay amplitude of $ X(2370)(p_a)/X(1835)(p_a)/\eta_c(p_a) \to
\eta(p_b)[\eta^\prime(p_b)] P(p_c)P(p_d)$ can be described in terms of the
Mandelstam variables
\begin{align}
s &= (p_c+p_d)^2 , \qquad
t=(p_a-p_c)^2\,, \qquad u=(p_a-p_d)^2\,,\nn\\
3s_0&\equiv
s+t+u=
 m_{\text{initial}}^2+m_{\eta(\eta^\prime)}^2+2m_P^2  \,.
\end{align}

The helicity angle $\theta$ is defined as the angle between one $\pi (K)$ and the $X(2370)/X(1835)/\eta_c$ in the rest frame of the $\pi\pi (K\bar{K})$ system.
$t$ and $u$ can be expressed in terms of $s$ and $\theta$ according to
\begin{align}
t &= \frac{1}{2} \left[3s_0-s+\kappa(s)\cos\theta \right]\,,&
u &= \frac{1}{2} \left[3s_0-s-\kappa(s)\cos\theta \right]\,, \nn\\
\kappa(s) &\equiv \sigma_P \lambda^{1/2}\big(m_{\text{initial}}^2,m_{\eta(\eta^\prime)}^2,s\big) \,, &
\sigma_P &\equiv \sqrt{1-\frac{4m_P^2}{s}} \,,
\label{eq:tu}
\end{align}
where $\lambda(a,b,c)=a^2+b^2+c^2-2(ab+ac+bc)$.

Using the Lagrangians in
Eqs.~\eqref{LagrangianXPPP} and~\eqref{LagrangianetacPPP}, we can obtain the projections of the $S$- and $D$-waves of the chiral contact terms of $ X(2370)/X(1835)/\eta_c \to \eta[\eta^\prime] \pi\pi(K\bar{K}) $ processes. 
In order not to interrupt the present discussion, the lengthy expressions are relegated to Appendix~\ref{supp:amplitudesSDwave}.

\subsection{Final-state interactions with a dispersive approach, Omn\`es solution }

There are strong final-state interactions in the $\pi\pi$ and $K\bar{K}$ systems. In this work, we will use dispersion theory to take into account the coupled-channel ($\pi\pi$ and $K\bar K$) FSI for the dominant $S$- and $D$-wave components. 
Similar methods to consider the FSI have been used previously e.g.\ in
Refs.~\cite{Anisovich:1996tx,Garcia-Martin:2010kyn,Kubis:2015sga,Dai:2014zta,Yao:2020bxx,Chen:2016mjn,Chen:2019gty,Chen:2019mgp,Chen:2021aud,Chen:2023def,Chen:2026fnz}.

For $ X(2370)/X(1835)/\eta_c \to \eta[\eta^\prime] \pi\pi(K\bar{K}) $, the
partial-wave decomposition of the amplitude including the $s$-channel FSI reads
\be
M^{R, \eta(\eta^\prime)PP,\text{decay}}(s,\cos\theta)  =\sum_{l=0}^{\infty} M_l^{R,\eta(\eta^\prime)PP}(s) P_l(\cos\theta)\,,
\label{eq.PartialWaveFullAmplitude}\ee
where ``$R$'' denotes different initial state $X(2370)/X(1835)/\eta_c$. $l$ represents the relative orbital angular momentum of
the pions (kaons), and we only consider the $S$-and $D$-wave components and therefore $l = 0, 2$.

The two-channel unitarity condition reads
\begin{equation}\label{eq.unitarity2channel}
\textrm{disc}\, \vec{M}_l^R(s)=2i T_l^{0\ast}(s)\Sigma(s)
\vec{M}_l^R(s),
\end{equation}
where the two-dimensional vectors $\vec{M}_l^R(s)$ contains both the $\pi\pi$ and the $K\bar{K}$ final states,
 \begin{equation}
\vec{M}_l^R(s)=\left( {\begin{array}{*{2}c}
   {M_l^{R, \eta(\eta^\prime)\pi\pi}(s)} \\
   {\frac{2}{\sqrt{3}}M_l^{R, \eta(\eta^\prime)K\bar{K}}(s)}  \\\end{array}} \right).
 \end{equation}
The two-dimensional matrices $T_l^0(s)$ and $\Sigma(s)$ are represented as
\begin{equation}\label{eq.Tl0}
T_l^0(s)=
 \left( {\begin{array}{*{2}c}
   \frac{\eta_l^0(s)e^{2i\delta_l^0(s)}-1}{2i\sigma_\pi(s)} & |g_l^0(s)|e^{i\psi_l^0(s)}   \\
  |g_l^0(s)|e^{i\psi_l^0(s)} & \frac{\eta_l^0(s)e^{2i\left(\psi_l^0(s)-\delta_l^0(s)\right)}-1}{2i\sigma_K(s)} \\
\end{array}} \right),
\end{equation}
and $\Sigma(s)\equiv \text{diag}
\big(\sigma_\pi(s)\theta(s-4m_\pi^2),\sigma_K(s)\theta(s-4m_K^2)\big)$.
There are three input functions in the $T_l^0(s)$ matrix: the
$\pi\pi$ $S/D$-wave isoscalar phase shift $\delta_l^0(s)$, and the modulus and phase of the $\pi\pi \to
K\bar{K}$ $S/D$-wave amplitude $g_l^0(s)=|g_l^0(s)|e^{i\psi_l^0(s)}$. 
Note that the relation between the  inelasticity parameter $\eta_l^0(s)$ in Eq.~\eqref{eq.Tl0} and the modulus $|g_l^0(s)|$
\begin{equation}
\eta_l^0(s)=\sqrt{1-4\sigma_\pi(s)\sigma_K(s)|g_l^0(s)|^2\theta(s-4m_K^2)}\,.
\end{equation}

\noindent The solution of the coupled-channel unitarity condition in
Eq.~\eqref{eq.unitarity2channel} is given by
\begin{equation}\label{OmnesSolution2channel}
\vec{M}_l^R(s)=\Omega_l(s)\vec{P}_l^{n-1}(s)
\,,
\end{equation}
where $\Omega_l(s)$ satisfies the homogeneous coupled-channel
unitarity relation
\begin{equation}\label{eq.unitarity2channelhomo}
\textrm{Im}\, \Omega_l(s)=T_l^{0\ast}(s)\Sigma(s) \Omega_l(s),
\hspace{1cm}  \Omega_l(0)=\mathbbm{1} \,.
\end{equation}

For the $S$-wave, the values of the input functions in the $T_0^0(s)$ matrix are taken from~\cite{Dai:2014lza,Dai:2014zta}. 
These inputs are used up to $\sqrt{s_0}=1.3\GeV$, and above $s_0$ the $f_0(1370)$ and $f_0(1500)$ resonances coupling strongly to
$4\pi$ will contribute additional inelasticities~\cite{ParticleDataGroup:2026mpi,Ropertz:2018stk}. Above $s_0$, we guide smoothly the phases
$\delta_0^0(s)$ and $\psi_0^0$ to 2$\pi$~\cite{Moussallam:1999aq},
\begin{equation}
\delta(s)=2\pi+(\delta(s_0)-2\pi)\frac{2}{1+({s}/{s_0})^{3/2}}\,.
\end{equation}
The numerical results of $\Omega_0(s)$ have been computed, e.g., in
Refs.~\cite{Moussallam:1999aq,Hoferichter:2012wf,Daub:2015xja}.
For the $D$-wave, we will use the result of $\Omega_2(s)$ given in Ref.~\cite{Danilkin:2025kyo}, in which the $f_2(1270)$ resonance in the $\pi\pi$ channel and the $f_2^\prime(1525)$ resonance in the $K\bar{K}$ channel are considered automatically.

On the other hand, at low energies the partial-wave amplitudes $\vec{M}_l^R(s)$ should match to those from $\chi$EFT. Namely, if one switches off the FSI at $s=0$, $\Omega_l(0)=\mathbbm{1}$,
the polynomial $\vec{P}_l^{n-1}(s)$ in Eq~\eqref{OmnesSolution2channel} should agree well with the low-energy chiral amplitudes. 
Thus, the amplitudes take the form
\begin{equation}\label{M02channel}
\vec{M}_l^R(s)=\Omega_l(s)\vec{M}_l^{R,\chi}(s)
\,,
\end{equation}
where $ \vec{M}_l^{R,\chi}(s)=\big(
   M_l^{R,\chi,\eta(\eta^\prime)\pi\pi}(s),
   2/\sqrt{3}\,M_l^{R,\chi,\eta(\eta^\prime)K\bar{K}}(s)
   \big)^{T}$, with the expressions of $M_l^{R,\chi,\eta(\eta^\prime)\pi\pi(K\bar{K})}(s)$ given in Eqs.~\eqref{AmpetapKK0}-\eqref{Ampetapipi2}.

The amplitude of the $ J/\psi \to \gamma\eta[\eta^\prime] \pi\pi(K\bar{K}) $  can be written as
\begin{eqnarray}
\mathscr{M}^{\eta(\eta^\prime)PP}(Q^2,s,\cos\theta)=&&2\epsilon^{\mu\nu\alpha\beta}p_\psi^\mu p_\gamma^\alpha \epsilon_\psi^\nu \epsilon_\gamma^\beta \bigg\{ \frac{g_\gamma^{X(1835)}}{Q^2-M_{X(1835)}^2+iM_{X(1835)}\Gamma_{X(1835)}}M^{X(1835), \eta(\eta^\prime)PP,\text{decay}}(s,\cos\theta) \nn\\ && +\frac{g_\gamma^{X(2370)}}{Q^2-M_{X(2370)}^2+iM_{X(2370)}\Gamma_{X(2370)}}M^{X(2370), \eta(\eta^\prime)PP,\text{decay}}(s,\cos\theta)+\frac{g_\gamma^{\eta_c}}{Q^2-M_{\eta_c}^2+iM_{\eta_c}\Gamma_{\eta_c}}  \nn\\ && M^{\eta_c, \eta(\eta^\prime)PP,\text{decay}}(s,\cos\theta) + M_{b.g.}^{\eta(\eta^\prime)}\bigg\},\label{eq.JpsitogammaPPPAmplitude}
\end{eqnarray}
where $Q$ is the invariant mass of the
$\eta(\eta^\prime)PP$ system, and we set the  $J/\psi  \gamma X(2370)/X(1835)/\eta_c$ coupling
constants as $g_\gamma^{X(1835)}=g_\gamma^{X(2370)}=g_\gamma^{\eta_c} =1$ since they can be absorbed into the coupling constants in the Lagrangians of $X(2370)/X(1835)/\eta_c \to PPP$ transitions in the fitting of the data. Here we use the energy-independent widths for the
$X(1835)$, $X(2370)$, and $\eta_c$. The values of the $X(1835)$ mass and width are taken as $1844\MeV$ and $192\MeV$, respectively, from Ref~\cite{BESIII:2015xco}. The mass and width of the $X(2370)$ are taken as $2359\MeV$ and $170\MeV$, respectively, from Ref~\cite{BESIII:2026rzt}. 
The mass and width of the $\eta_c$ are taken from PDG~\cite{ParticleDataGroup:2026mpi}. $M_{b.g.}$ is a background [not from the intermediate states $X(1835)$, $X(2370)$, and $\eta_c$] parametrized by
first-order polynomials of $Q^2$: $M_{b.g.} = \beta_0^{\eta(\eta^\prime)}+ \beta_1^{\eta(\eta^\prime)} Q^2$.

The differential decay width for the process $J/\psi \to \gamma\eta(\eta^\prime)PP $ can be written as
\bea
\frac{\diff^3 \Gamma}{\diff Q \diff s \diff t}&=&
N^{\eta(\eta^\prime)PP}\frac{(M_\psi^2-Q^2) }{128Q M_\psi^3(2\pi)^5}  \left|\mathscr{M}^{\eta(\eta^\prime)PP}(Q^2,s,\cos\theta)\right|^2\,,
\eea
where $N^{\eta(\eta^\prime)PP}$ are the normalization factors since the experimental data are given in events.
The limits of the varables are
\bea
Q_-&=&m_{\eta(\eta^\prime)}+2m_P\;,\nonumber\\
Q_+&=&M_{\psi}\;,\nonumber\\
s_-&=&4m_P^2\;,\nonumber\\
s_+&=&(Q-m_{\eta(\eta^\prime)})^2\;,\nonumber\\
t_{\pm}&=&\frac{1}{4s}\left( (Q^2-m_{\eta(\eta^\prime)}^2)^2-[\lambda^{1/2}(Q^2,s,m_{\eta(\eta^\prime)}^2) \mp \lambda^{1/2}(s,m_P^2,m_P^2)]^2 \right)\,.
\eea
Using the relation between $t$ and $\cos\theta$ in Eq.~\eqref{eq:tu} one also can get the helicity angular distribution of the $J/\psi \to \gamma\eta(\eta^\prime)PP $ decays.

The differential decay width for $ X(2370)  \to \eta(\eta^\prime) PP $ reads
\begin{equation}\label{eq.DecayWidthX2370toPPP}
\frac{\diff\Gamma}{\diff \sqrt{s} \,\diff\cos\theta} =
\frac{\sqrt{s}\,\sigma_P |\vec{q}|}{128\pi^3 M_{X(2370)}^2}
\left|M^{X(2370), \eta(\eta^\prime)PP,\text{decay}}(s,\cos\theta) \right|^2\,,
\end{equation}
where the limits of the varables are $s_-=4m_P^2$ and $s_+=(M_{X(2370)}-m_{\eta(\eta^\prime)})^2$, and
$\vec{q}$ is the 3-momentum of the final $\eta(\eta^\prime)$ in the rest frame of the initial $ X(2370)$ with
\be \label{eq:q}
|\vec{q}|=\frac{1}{2M_{X(2370)}}
\lambda^{1/2}\big(M_{X(2370)}^2,m_{\eta(\eta^\prime)}^2,s\big) \,.
\ee

 \end{onecolumngrid}

\medskip

\section{Phenomenological discussion}\label{results}

In this work we perform fits taking into account the experimental data sets of the $K_S^0 K_S^0 \eta$, $K_S^0 K_S^0$, and $K_S^0 \eta$ invariant mass distributions and the angular distribution of $K_S^0$ in the $K_S^0 K_S^0$ rest system of $J/\psi \to \gamma K_S^0 K_S^0 \eta$ process~\cite{BESIII:2015xco}, the $K_S^0 K_S^0 \eta^\prime$, $K_S^0 K_S^0$, and $K_S^0 \eta^\prime$ invariant mass distributions and the angular distribution of $K_S^0$ in the $K_S^0 K_S^0$ rest system of $J/\psi \to \gamma K_S^0 K_S^0 \eta^\prime$ process~\cite{BESIII:2023wfi}, and the decay-width ratio of 
$\Gamma(X(2370) \to K\bar{K}\eta^\prime)/\Gamma(X(2370) \to \pi\pi\eta^\prime)=\Gamma(J/\psi \to \gamma X(2370) \to \gamma K\bar{K}\eta^\prime)/\Gamma(J/\psi \to \gamma X(2370) \to \gamma \pi\pi\eta^\prime)=0.201\pm 0.108$~\cite{BESIII:2026mvn}. Note that to reduce the additional intermediate processes, the events satisfying $M_{K_S^0 K_S^0} < 1.1$ GeV$/c^2$ are used in Refs~\cite{BESIII:2015xco,BESIII:2023wfi}. In our fits we do not use the product branching fractions of $J/\psi \to \gamma X(2370) \to \gamma K\bar{K}\pi$ and $J/\psi \to \gamma X(2370) \to \gamma \pi\pi\eta$ given in Ref.~\cite{BESIII:2026mvn}, since in this study we do not consider the mechanism of $X(2370) \to K_0^\ast(1430)K \to K\bar{K}\pi$ and the effect of $f_0(1500)$ which may play important roles in these two processes, respectively. 

Using the constraint $g_{8b}^X=h_8^X g_1^X/h_1^X-2g_{8a}^X$,\footnote{One can construct the Lagrangian using Eq.~\eqref{eq.YComponents} as the interpolating field for the $X(2370)/X(1835)$ directly, which is equivalent to writing the Lagrangian in the form of Eq.~\eqref{LagrangianXPPP} with $(2g_{8a}^X+g_{8b}^X)/g_1^X = h_8^X/h_1^X = b^X/a^X.$} there are 16 free parameters in our fits: $g_1^{X(2370)/X(1835)/\eta_c}$, $h_1^{X(2370)/X(1835)/\eta_c}$, $g_{8a}^{X(2370)/X(1835)}$, $h_{8}^{X(2370)/X(1835)}$, $\beta_{0(1)}^{\eta(\eta^\prime)}$, and two normalization factors $N^{\eta(\eta^\prime)PP}$. The parameters  $g_1^{X(2370)/X(1835)/\eta_c}$ and $h_1^{X(2370)/X(1835)/\eta_c}$
correspond to the low-energy constants in the $XPPP$ and $\eta_c PPP$ Lagrangians in Eqs.~\eqref{LagrangianXPPP} and~\eqref{LagrangianetacPPP} for the SU(3) singlet component of the $X(2370)/X(1835)/\eta_c$, $g_{8a}^{X(2370)/X(1835)}$ and $h_{8}^{X(2370)/X(1835)}$ are the corresponding parameters for the SU(3) octet component of $X(2370)/X(1835)$. $\beta_{0(1)}^{\eta(\eta^\prime)}$ are the paramters in the background defined below Eq.~\eqref{eq.JpsitogammaPPPAmplitude}. 
To illustrate the effect of the SU(3) octet components of $X(1835)$ and $X(2370)$, we perform four fits. To be specific, in Fit I we only consider the SU(3) singlet components for $X(1835)$ and $X(2370)$;
 in Fit~II, both the SU(3) singlet and octet components are taken into account for $X(1835)$, while only the SU(3) singlet component is considered for $X(2370)$;
 in Fit~III, only the SU(3) singlet component is taken into account for $X(2370)$, while both the SU(3) singlet and octet components are considered for $X(1835)$;
  in Fit~IV, we consider both the SU(3) singlet and octet components for $X(1835)$ and $X(2370)$.
The coupled-channel FSI is considered in all four fits.

\begin{figure}
\centering
\includegraphics[width=\linewidth]{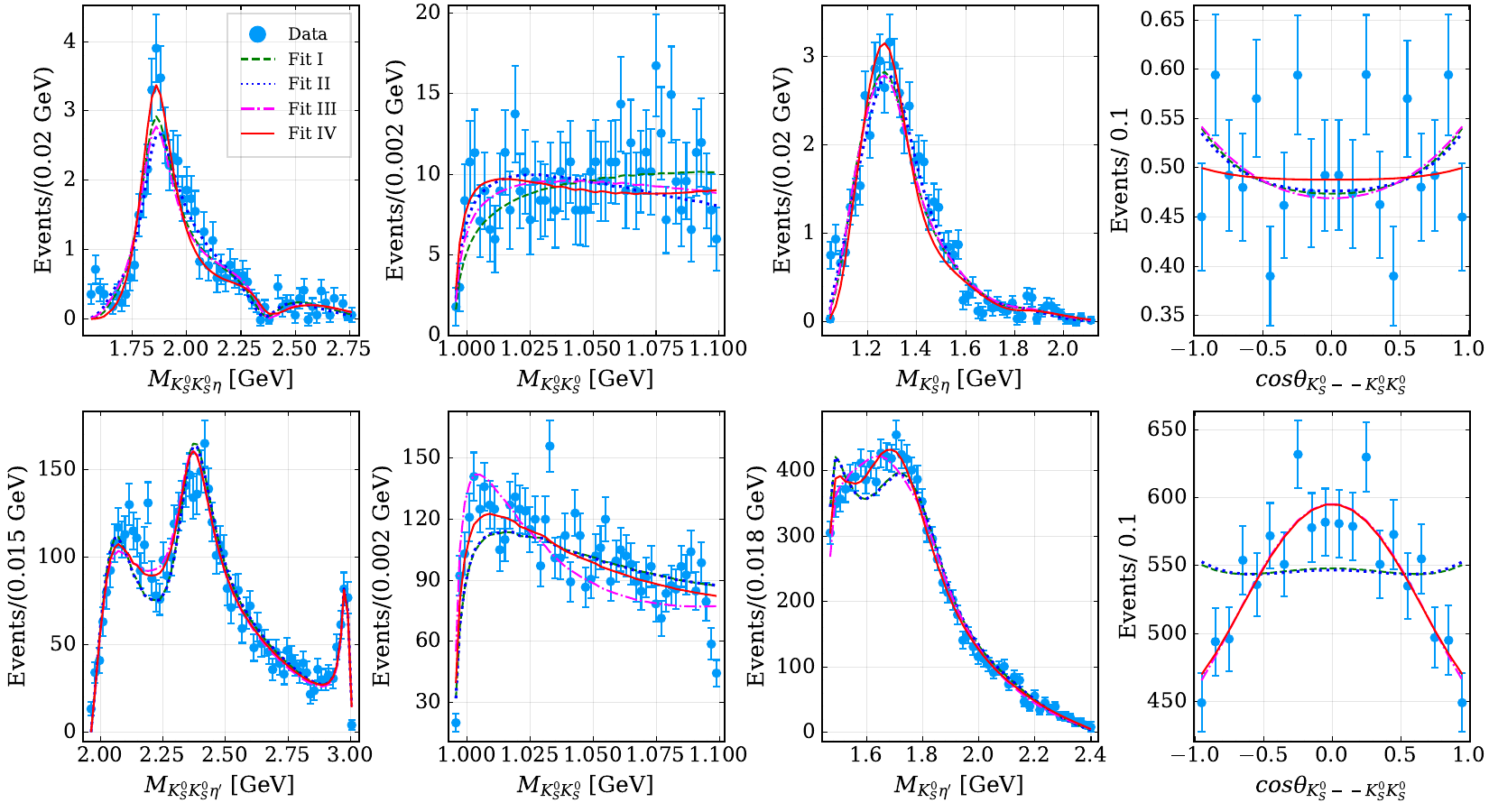}
\caption{Fit results of the $K_S^0 K_S^0 \eta$, $K_S^0 K_S^0$, and $K_S^0 \eta$ invariant mass distributions and the angular distribution of $K_S^0$ in the $K_S^0 K_S^0$ rest system of $J/\psi \to \gamma K_S^0 K_S^0 \eta$ process (top panel, from left to right), the $K_S^0 K_S^0 \eta^\prime$, $K_S^0 K_S^0$, and $K_S^0 \eta^\prime$ invariant mass distributions and the angular distribution of $K_S^0$ in the $K_S^0 K_S^0$ rest system of $J/\psi \to \gamma K_S^0 K_S^0 \eta^\prime$ process (bottom panel, from left to right). The results of Fits I, II, III, and IV are shown as the green dashed, blue dotted, magenta dot-dashed, and red solid lines, respectively. The experimental data are taken from Refs.~\cite{BESIII:2015xco} and~\cite{BESIII:2023wfi}. }\label{fig.fitresults}
\end{figure}

\begin{table}
\caption{\label{table-ratioes}  Experimental and theoretical
values  for the decay-width ratio of 
$\Gamma(X(2370) \to K\bar{K}\eta^\prime)/\Gamma(X(2370) \to \pi\pi\eta^\prime)=\Gamma(J/\psi \to \gamma X(2370) \to \gamma K\bar{K}\eta^\prime)/\Gamma(J/\psi \to \gamma X(2370) \to \gamma \pi\pi\eta^\prime)$~\cite{BESIII:2026mvn}. The experimental data are taken
from Ref.~\cite{BESIII:2026mvn}. }
\renewcommand{\arraystretch}{1.2}
\begin{center}
\begin{tabular}{l|ccccc}
\toprule
&Experiment &           Fit~I &   Fit~II &           Fit~III &   Fit~IV  \\
\hline
 $\frac{\Gamma(X(2370) \to K\bar{K}\eta^\prime)}{\Gamma(X(2370) \to \pi\pi\eta^\prime)} $ &  $0.201\pm 0.108$ & $0.306\pm 0.014$& $0.305\pm 0.032$ &   $0.251\pm 0.081$& $0.163\pm 0.062$\\
\botrule
\end{tabular}
\end{center}
\renewcommand{\arraystretch}{1.0}
\end{table}

\begin{table}
\caption{\label{tablepar1} Fit parameters from the simultaneous fit of the $K_S^0 K_S^0 \eta$, $K_S^0 K_S^0$, and $K_S^0 \eta$ invariant mass distributions and the angular distribution of $K_S^0$ in the $K_S^0 K_S^0$ rest system of $J/\psi \to \gamma K_S^0 K_S^0 \eta$ process, the $K_S^0 K_S^0 \eta^\prime$, $K_S^0 K_S^0$, and $K_S^0 \eta^\prime$ invariant mass distributions and the angular distribution of $K_S^0$ in the $K_S^0 K_S^0$ rest system of $J/\psi \to \gamma K_S^0 K_S^0 \eta^\prime$ process, and the decay-width ratio of 
$\Gamma(X(2370) \to K\bar{K}\eta^\prime)/\Gamma(X(2370) \to \pi\pi\eta^\prime)=\Gamma(J/\psi \to \gamma X(2370) \to \gamma K\bar{K}\eta^\prime)/\Gamma(J/\psi \to \gamma X(2370) \to \gamma \pi\pi\eta^\prime)$. Fit I: considering only the SU(3) singlet components for $X(1835)$ and $X(2370)$; Fit II: considering both the SU(3) singlet and octet components for $X(1835)$, and considering only the SU(3) singlet component for $X(2370)$; Fit III: considering only the SU(3) singlet component for $X(1835)$, and considering both the SU(3) singlet and octet components for $X(2370)$; Fit IV: considering both the SU(3) singlet and octet components for $X(1835)$ and $X(2370)$. }
\renewcommand{\arraystretch}{1.2}
\begin{center}
\begin{tabular}{l|cccc}
\toprule
         & Fit~I    & Fit~II & Fit~III       & Fit~IV\\
\hline
$g_1^{X(1835)}~[\text{GeV}^{-1}]$   &    $ -8.29 \pm	0.24$  &   $ -7.6 \pm 	0.4$  &    $-4.68 \pm	0.33$  &   $ -8.29\pm 0.16$\\
$h_1^{X(1835)}~[\text{GeV}^{-1}]$   &    $ 31.0 \pm	0.7$  &   $ 20.1 \pm	0.8$  &    $ 15.0 \pm	0.4$  &   $16.08 \pm 0.31$ \\
$g_{8a}^{X(1835)}~[\text{GeV}^{-1}]$     &    0 (\text{fixed})   & $-3.29\pm	0.10$ &    0 (\text{fixed})   &   $ 6.35 \pm	0.33$ \\
$h_8^{X(1835)}~[\text{GeV}^{-1}]$     &    0 (\text{fixed})   & $-5.5	\pm 0.1$  &   0 (\text{fixed})   &   $ -9.5 \pm	0.7$ \\
$g_1^{X(2370)}~[\text{GeV}^{-1}]$   &    $ 3.07 \pm	0.05$  &   $ 2.16\pm	0.08$  &    $ 8.93 \pm	0.16$  &   $ 7.91	\pm 0.12$\\
$h_1^{X(2370)}~[\text{GeV}^{-1}]$   &    $ -22.59	\pm 0.28$  &   $ -15.8 \pm	0.5$  &    $ -37.2 \pm	0.5 $  &   $ -32.69\pm	0.25$ \\
$g_{8a}^{X(2370)}~[\text{GeV}^{-1}]$     &    0 (\text{fixed})   &  0 (\text{fixed})   &    $ -4.43	\pm 0.06$  &   $ -3.68	\pm 0.05$\\
$h_8^{X(2370)}~[\text{GeV}^{-1}]$     &    0 (\text{fixed})   &  0 (\text{fixed})   &    $ 46.7 \pm	0.4$  &   $ 39.68 \pm	0.14$ \\
$g_1^{\eta_c}~[\text{GeV}^{-1}]$   &    $ 2.61 \pm	0.12$  &   $ 1.75	\pm 0.14$  &    $ 0.57 \pm	0.19$  &   $ 0.50	\pm 0.15$\\
$h_1^{\eta_c}~[\text{GeV}^{-1}]$   &    $ -3.2 \pm	1.3$  &   $ -1.6	\pm 1.3$  &    $ -17.2 \pm	1.6$  &   $ -14.4\pm	1.0$ \\
$\beta_{0}^{\eta} \times 10^{-3}$  &    $ 52.3\pm 2.5$    &   $ -56.0\pm 7.1$  &    $ 19.0\pm 7.1$  &   $ 131.0\pm 11.5$\\
$\beta_{1}^{\eta} \times 10^{-3}$  &    $ -46.9\pm 1.2$    &   $ 16.8\pm 3.4 $ &    $-14.2\pm 3.3$  &   $ -81.0\pm 5.1$ \\
$\beta_{0}^{\eta^\prime} \times 10^{-3}$  &    $ 27.0\pm 2.4$    &   $ 23.0\pm 7.1$ &    $ -14.0\pm 3.3$  &   $ -47.4\pm 1.2$ \\
$\beta_{1}^{\eta^\prime} \times 10^{-3}$  &    $ -21.6\pm 1.0$    &   $ -15.6\pm 2.8$ &    $ -11.1\pm 2.0$  &   $ 13.8\pm 0.4$ \\
\hline
 ${\chi^2}/{\rm d.o.f.}$ &  $\frac{796.3}{(384-12)}=2.14$  &  $\frac{773.3}{(384-14)}=2.09$  &  $\frac{682.0}{(384-14)}=1.84$  &  $\frac{599.9}{(384-16)}=1.63$  \\
\botrule
\end{tabular}
\end{center}
\renewcommand{\arraystretch}{1.0}
\end{table}

In Fig.~\ref{fig.fitresults}, the fitted results of the invariant mass distributions and the angular distribution in Fits I, II, III, and IV are shown as the green dashed, blue dotted, magenta dot-dashed, and red solid lines, respectively. The fit results of the decay-width ratio of $\Gamma(X(2370) \to K\bar{K}\eta^\prime)/\Gamma(X(2370) \to \pi\pi\eta^\prime)$ are given in Table~\ref{table-ratioes}. The fitted parameters as well as the $\chi^2/\text{d.o.f.}$ are shown in Table~\ref{tablepar1}.

As can be seen from Fig.~\ref{fig.fitresults}, in the $K_S^0 K_S^0 \eta$ invariant mass distribution the $X(1835)$ is dominant, while the $X(2370)$ and the interference effect account for the dip around 2.37 GeV.\footnote{The $X(1560)$ state announced in~\cite{BESIII:2015xco} contributing the small peak near the threshold is not considered here, since its effect is tiny and it will introduce additional free parameters.} In the $K_S^0 K_S^0 \eta^\prime$ invariant mass distribution, the $X(2370)$ peak is the most prominent.

It is interesting to compare these four fits. 
First, one observes that in Fits I and II only considering the SU(3) singlet components for $X(2370)$, the hump near the threshold in the $K_S^0 K_S^0$ invariant mass distributions and the angular distribution of $J/\psi \to \gamma K_S^0 K_S^0 \eta^\prime$ process cannot be well described well. This indicates the necessity of taking into account the SU(3) octet component for $X(2370)$. In contrast, in Fits III and IV, including the SU(3) octet component for $X(2370)$ in addition, the fit qualities are improved significantly. Note that the experimental data of the angular distribution of $K_S^0$ in the $K_S^0 K_S^0$ rest system of $J/\psi \to \gamma K_S^0 K_S^0 \eta^\prime$ process is not flat, which means the $D$-wave contribution can not been neglected. As given in the Lagrangian in Eq.~\eqref{LagrangianXPPP}, the $h_{1,(8)}^{X(2370)}$ terms are energy-independent and contribute to pure $S$ wave, and the $g_{1,(8a,8b)}^{X(2370)}$ terms are energy-dependent and contribute to both $S$ and $D$ waves. For the fitted results of the parameters presented in Table~\ref{tablepar1}, one finds that in Fits III and IV the values of $g_{1,(8a)}^{X(2370)}$ are one order of magnitude smaller than the values of $h_{1,(8)}^{X(2370)}$. The dominance of the energy-independent terms agrees with our expectations, while the energy-dependent terms are indispensable to reproduce the angular distribution.
Second, comparing Fits III and IV, one finds that including the SU(3) octet component for $X(1835)$ mainly leads the peak in the $K_S^0 K_S^0 \eta$ invariant mass distribution somewhat higher, while in other places the differences are small. This indicates that the effect of the octet component for $X(1835)$ is not significant.

Using the best fit results of Fit~IV given in Tables~\ref{tablepar1}, we can calculate the ratio of the parameter for the SU(3) octet component relative to that for
the SU(3) singlet component,  $h_8^{X(2370)}/h_1^{X(2370)}=-1.21 \pm 0.01$. Thus we conclude that there is a large
light-quark SU(3) octet component in the $X(2370)$, and the scenario of a glueball state is disfavored since the light quarks have to be produced in the flavor-singlet state in such state.

\begin{figure}
\centering
\includegraphics[width=\linewidth]{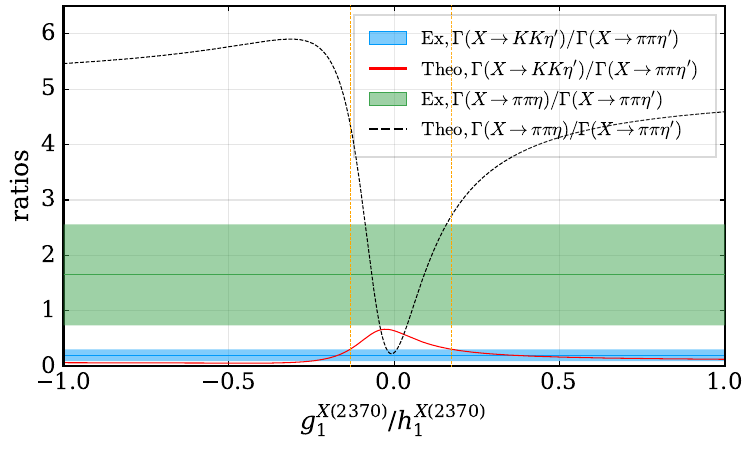}
\caption{Experimental and theoretical values for the decay widths ratios ${\Gamma(X(2370) \to K\bar{K}\eta^\prime)}/{\Gamma(X(2370) \to \pi\pi\eta^\prime)}$ and ${\Gamma(X(2370) \to \pi\pi\eta)}/{\Gamma(X(2370) \to \pi\pi\eta^\prime)}$. The theoretical results are calculated considering only the singlet terms for $X(2370)$ with varying $g_1^{X(2370)}/h_1^{X(2370)}$
in the range of $(-1, 1)$.  The experimental data are extracted from Ref.~\cite{BESIII:2026mvn}. The two dashed vertical lines correspond to the two points where the theoretical values for the decay widths ratios ${\Gamma(X(2370) \to K\bar{K}\eta^\prime)}/{\Gamma(X(2370) \to \pi\pi\eta^\prime)}$ attach the upper edges of corresponding experimental error bar.}\label{fig.PredictionOfRatios}
\end{figure}

It is instructive to discuss what would be expected for the decay widths ratios produced from pure flavor singlet of $X(2370)$. In Fig.~\ref{fig.PredictionOfRatios}, we plot the theoretical values for the decay widths ratios ${\Gamma(X(2370) \to K\bar{K}\eta^\prime)}/{\Gamma(X(2370) \to \pi\pi\eta^\prime)}$ and ${\Gamma(X(2370) \to \pi\pi\eta)}/{\Gamma(X(2370) \to \pi\pi\eta^\prime)}$, which are calculated considering only the flavor singlet terms for $X(2370)$ with varying $g_1^{X(2370)}/h_1^{X(2370)}$ in the range of $(-1, 1)$. For comparison, the experimental values extracted from Ref.~\cite{BESIII:2026mvn} are also shown. The two dashed vertical lines correspond to the two points where the theoretical values for the ratio ${\Gamma(X(2370) \to K\bar{K}\eta^\prime)}/{\Gamma(X(2370) \to \pi\pi\eta^\prime)}$ attach the upper edges of corresponding experimental error bar. First, one observes that there is no region of $g_1^{X(2370)}/h_1^{X(2370)}$ at which the theoretical values for the decay widths ratios ${\Gamma(X(2370) \to K\bar{K}\eta^\prime)}/{\Gamma(X(2370) \to \pi\pi\eta^\prime)}$ and ${\Gamma(X(2370) \to \pi\pi\eta)}/{\Gamma(X(2370) \to \pi\pi\eta^\prime)}$ can lie within corresponding experimental error bars simultaneously. Note that in our theoretical scheme we do not consider the effect of $f_0(1500)$ to avoid introducing more free parameters. The effect of $f_0(1500)$ in $X(2370) \to \pi\pi \eta^\prime (K\bar{K}\eta^\prime)$ processes should be suppressed by the small phase space. While in $X(2370) \to \pi\pi \eta$ process, taking account of the effect of $f_0(1500)$ is expected to enhance the product of the $\pi\pi \eta$ events and therefore enhance the theoretical plot of ${\Gamma(X(2370) \to \pi\pi\eta)}/{\Gamma(X(2370) \to \pi\pi\eta^\prime)}$ in Fig.~\ref{fig.PredictionOfRatios}, since in general $f_0(1500)$ contributes constructive interference with $f_0(980)$ as shown, for example, in the $B_s^0 \to J/\psi\pi\pi (\psi(2S)\pi\pi)$ processes~\cite{LHCb:2014ooi,LHCb:2023reb}. As can be inferred from Fig.~\ref{fig.PredictionOfRatios}, considering the effect of $f_0(1500)$ will make it even more difficult for the theoretical values of the decay widths ratios ${\Gamma(X(2370) \to K\bar{K}\eta^\prime)}/{\Gamma(X(2370) \to \pi\pi\eta^\prime)}$ and ${\Gamma(X(2370) \to \pi\pi\eta)}/{\Gamma(X(2370) \to \pi\pi\eta^\prime)}$ to be  both in agreement with corresponding experimental values. Therefore, we conclude that if only consider the flavor singlet of $X(2370)$, the decay widths ratios ${\Gamma(X(2370) \to K\bar{K}\eta^\prime)}/{\Gamma(X(2370) \to \pi\pi\eta^\prime)}$ and ${\Gamma(X(2370) \to \pi\pi\eta)}/{\Gamma(X(2370) \to \pi\pi\eta^\prime)}$ can not be described simultaneously. In contrast, we have tried performing new fits also including the decay widths ratio ${\Gamma(X(2370) \to \pi\pi\eta)}/{\Gamma(X(2370) \to \pi\pi\eta^\prime)}$ in the treatment taking account of both the singlet and octet components of $X(2370)$. We find that all the data can be described well, while we don't exhibit these fit results here since the effect of $f_0(1500)$ is not considered in the present scheme. Second, notice that in Refs.~\cite{Eshraim:2012jv,Eshraim:2020zrs,Giacosa:2026pqh}, the $ X(2370) \to PPP $ decay widths have been calculated using a method based the extended linear sigma model~\cite{Parganlija:2016yxq}, in which only energy-independent couplings are considered. In their scheme, the $X(2370)$ is treated as a glueball and thus as a flavor singlet, and the seuential process $X(2370) \to PS \to P\pi\pi (P KK)$, where the scalar $S$ can correspond to the $f_0(500)$ and the $f_0(980)$, has been taken into account. In our scheme, by using the inputs from known scattering observables in the dispersion relations, the effects of resonances in the considered partial waves, i.e., the $f_0(500)$, $f_0(980)$, $f_2(1270)$, and $f_2^\prime(1525)$, are included
automatically in a model-independent way. From Fig.~\ref{fig.PredictionOfRatios}, one observes that at $g_1^{X(2370)}/h_1^{X(2370)}=0$, the theoretical value of the ratio ${\Gamma(X(2370) \to K\bar{K}\eta^\prime)}/{\Gamma(X(2370) \to \pi\pi\eta^\prime)}$ deviates from the experimental value. As stated above, the $g_1^{X(2370)}$ term and the $h_1^{X(2370)}$ term are energy-dependent and energy-independent, respectively. Therefore, we would like to point out that if the $X(2370)$ is regarded as a flavor singlet, both the energy-dependent and energy-independent couplings should be considered in the study of $X(2370) \to PPP$ processes.

\medskip

\section{Summary}\label{summary}

We have used dispersion theory to study the processes
$J/\psi \to \gamma  X(2370)/X(1835)/\eta_c \to \gamma  K\bar{K}\eta^{(\prime)} (\gamma  \pi\pi\eta^{(\prime)})$. In particular, we have analyzed the roles of the
light-quark SU(3) singlet state and SU(3) octet state in these
transitions. The strong coupled-channel ($\pi\pi$ and $K\bar{K}$) FSI in
the $S$- and $D$-waves has been considered in a model-independent way.
Through fitting to the data of the $K_S^0 K_S^0 \eta^{(\prime)}$, $K_S^0 K_S^0$ and $K_S^0 \eta^{(\prime)}$ invariant mass distributions and the angular distribution of $K_S^0$ in the $K_S^0 K_S^0$ rest system of $J/\psi \to \gamma K_S^0 K_S^0 \eta^{(\prime)}$ processes and the decay-width ratio of $\Gamma(X(2370) \to K\bar{K}\eta^\prime)/\Gamma(X(2370) \to \pi\pi\eta^\prime)$, we find that the light-quark SU(3) octet state plays a significant role in the  $X(2370)  K\bar K \eta^{(\prime)}$ and $X(2370) \pi\pi \eta^{(\prime)}$ transitions, which indicates that the $X(2370)$ contains a large light-quark component. Furthermore, we find that if only consider the flavor singlet of $X(2370)$, the decay widths ratios ${\Gamma(X(2370) \to K\bar{K}\eta^\prime)}/{\Gamma(X(2370) \to \pi\pi\eta^\prime)}$ and ${\Gamma(X(2370) \to \pi\pi\eta)}/{\Gamma(X(2370) \to \pi\pi\eta^\prime)}$ can not be described simultaneously. Therefore we conclude that the $X(2370)$ is in all likelihood neither a glueball nor a hybrid. 
Also, we find that the energy-dependent couplings should not be neglected in the study of the $X(2370) \to PPP$ processes.

\medskip
\medskip
\medskip



\medskip
\medskip
\medskip

\begin{onecolumngrid}

\newpage

%
\begin{appendix}
%

\begin{center}
  \textbf{\large\normalfont\bfseries\boldmath  Nature of the $X(2370)$: A light-quark perspective } \\
\vspace{0.05in}      \label{supp}
{ \it \large Supplementary Material}\\
\vspace{0.05in}
\end{center}

\section{$S$- and $D$-waves of amplitudes of $ X(2370)/X(1835)/\eta_c \to \eta[\eta^\prime] \pi\pi(K\bar{K}) $ processes}
\label{supp:amplitudesSDwave}

 For latter convenience, we define several $a_i$ factors as follows
 \begin{eqnarray}
a_1 = && \frac{F}{\cos(\theta_0-\theta_8)}\left(\frac{1}{\sqrt{6}}\frac{\cos \theta_0}{F_8}-\frac{1}{\sqrt{3}}\frac{\sin \theta_8}{F_0}\right)
\,, \nn \\
a_2 = && \frac{F}{\cos(\theta_0-\theta_8)}\left(\frac{1}{\sqrt{6}}\frac{\sin \theta_0}{F_8}+\frac{1}{\sqrt{3}}\frac{\cos \theta_8}{F_0}\right)
\,, \nn \\
a_3 = && \frac{F}{\cos(\theta_0-\theta_8)}\left(-\frac{2}{\sqrt{6}}\frac{\cos \theta_0}{F_8}-\frac{1}{\sqrt{3}}\frac{\sin \theta_8}{F_0}\right)
\,, \nn \\
a_4 = && \frac{F}{\cos(\theta_0-\theta_8)}\left(-\frac{2}{\sqrt{6}}\frac{\sin \theta_0}{F_8}+\frac{1}{\sqrt{3}}\frac{\cos \theta_8}{F_0}\right)
\nn\,.
\end{eqnarray}

The projections of the $S$- and $D$-waves of the chiral contact amplitudes of $ X(2370)/X(1835)/\eta_c \to \eta[\eta^\prime] \pi\pi(K\bar{K}) $ processes are given below:
\begin{align}\label{AmpetapKK0}
M_0^{R, \chi, \eta^\prime K\bar{K}}(s)=&  -\frac{1}{6 \sqrt{3} F F_K^2 s}
   \big\{2 \sqrt{2} a_2 g_1^{R} m_{\eta^\prime}^4 m_K^2 + 2 \sqrt{2} a_4 g_1^{R} m_{\eta^\prime}^4 m_K^2 -      4 \sqrt{2} a_2 g_1^{R} m_{\eta^\prime}^2 M_{R}^2 m_K^2 - 
     4 \sqrt{2} a_4 g_1^{R} m_{\eta^\prime}^2 M_{R}^2 m_K^2  \nn\\ & +  4 a_2 (g_{8a}^{R} - g_{8b}^{R}) (m_{\eta^\prime}^2 - M_{R}^2)^2 m_K^2 -      2 a_4 (4 g_{8a}^{R} - g_{8b}^{R}) (m_{\eta^\prime}^2 - M_{R}^2)^2 m_K^2 +      2 \sqrt{2} a_2 g_1^{R} M_{R}^2 m_K^2 (M_{R}^2 - s) \nn\\ & +    2 \sqrt{2} a_4 g_1^{R} M_{R}^2 m_K^2 (M_{R}^2 - s) +      \sqrt{2} a_2 g_1^{R} m_{\eta^\prime}^4 s + \sqrt{2} a_4 g_1^{R} m_{\eta^\prime}^4 s +      2 a_2 g_{8a}^{R} m_{\eta^\prime}^4 s - 4 a_4 g_{8a}^{R} m_{\eta^\prime}^4 s - 2 a_2 g_{8b}^{R} m_{\eta^\prime}^4 s  \nn\\ & + 
     a_4 g_{8b}^{R} m_{\eta^\prime}^4 s - 2 \sqrt{2} a_2 g_1^{R} m_{\eta^\prime}^2 M_{R}^2 s -      2 \sqrt{2} a_4 g_1^{R} m_{\eta^\prime}^2 M_{R}^2 s - 
     4 a_2 g_{8a}^{R} m_{\eta^\prime}^2 M_{R}^2 s + 8 a_4 g_{8a}^{R} m_{\eta^\prime}^2 M_{R}^2 s  \nn\\ & + 
     4 a_2 g_{8b}^{R} m_{\eta^\prime}^2 M_{R}^2 s - 2 a_4 g_{8b}^{R} m_{\eta^\prime}^2 M_{R}^2 s +      2 a_2 g_{8a}^{R} M_{R}^4 s - 4 a_4 g_{8a}^{R} M_{R}^4 s - 
     2 a_2 g_{8b}^{R} M_{R}^4 s + a_4 g_{8b}^{R} M_{R}^4 s  \nn\\ & - 
          20 a_2 g_{8a}^{R} m_{\eta^\prime}^2 m_K^2 s + 4 a_4 g_{8a}^{R} m_{\eta^\prime}^2 m_K^2 s +      20 a_2 g_{8b}^{R} m_{\eta^\prime}^2 m_K^2 s - 28 a_4 g_{8b}^{R} m_{\eta^\prime}^2 m_K^2 s + 
     4 a_2 g_{8a}^{R} M_{R}^2 m_K^2 s  \nn\\ & + 28 a_4 g_{8a}^{R} M_{R}^2 m_K^2 s -      4 a_2 g_{8b}^{R} M_{R}^2 m_K^2 s + 20 a_4 g_{8b}^{R} M_{R}^2 m_K^2 s + 
     6 \sqrt{2} a_4 h_1^{R} (24 m_K^4 - 6 m_K^2 m_{\pi}^2 - m_{\pi}^4) s  \nn\\ &  -      24 a_2 h_8^{R} (2 m_K^4 + m_K^2 m_{\pi}^2 - m_{\pi}^4) s -      24 a_4 h_8^{R} (6 m_K^4 - 6 m_K^2 m_{\pi}^2 + m_{\pi}^4) s + 
     6 \sqrt{2} a_2 h_1^{R} (4 m_K^4 + 20 m_K^2 m_{\pi}^2 + m_{\pi}^4) s \nn\\ & +      \sqrt{2} a_2 g_1^{R} M_{R}^2 (M_{R}^2 - s) s +      \sqrt{2} a_4 g_1^{R} M_{R}^2 (M_{R}^2 - s) s - 
     14 \sqrt{2} a_2 g_1^{R} m_K^2 (M_{R}^2 - s) s  \nn\\ & -      14 \sqrt{2} a_4 g_1^{R} m_K^2 (M_{R}^2 - s) s -      10 a_4 g_{8a}^{R} m_{\eta^\prime}^2 s^2 + 14 a_2 (g_{8a}^{R} - g_{8b}^{R}) m_{\eta^\prime}^2 s^2 + 
     16 a_4 g_{8b}^{R} m_{\eta^\prime}^2 s^2 + 2 a_4 g_{8a}^{R} M_{R}^2 s^2 \nn\\ &  -      10 a_2 (g_{8a}^{R} - g_{8b}^{R}) M_{R}^2 s^2 - 14 a_4 g_{8b}^{R} M_{R}^2 s^2 -      20 a_4 g_{8a}^{R} m_K^2 s^2 - 8 a_2 (g_{8a}^{R} - g_{8b}^{R}) m_K^2 s^2 -      22 a_4 g_{8b}^{R} m_K^2 s^2  \nn\\ & + 5 \sqrt{2} a_2 g_1^{R} (M_{R}^2 - s) s^2 +      5 \sqrt{2} a_4 g_1^{R} (M_{R}^2 - s) s^2 + 2 a_4 g_{8a}^{R} s^3 + 
     8 a_2 (g_{8a}^{R} - g_{8b}^{R}) s^3 + 13 a_4 g_{8b}^{R} s^3  \nn\\ & -      2 \sqrt{2} a_2 g_1^{R} m_{\eta^\prime}^2 s (-4 m_K^2 + s) -      2 \sqrt{2} a_4 g_1^{R} m_{\eta^\prime}^2 s (-4 m_K^2 + s)\big\}\,,
\end{align}

\begin{align}\label{AmpetapKK2}
M_2^{R,\chi, \eta^\prime K\bar{K}}(s)=& -\frac{1}{6 \sqrt{3} F F_K^2 s}
   \big\{4 \sqrt{2} a_2 g_1^{R} m_{\eta^\prime}^4 m_K^2 +     4 \sqrt{2} a_4 g_1^{R} m_{\eta^\prime}^4 m_K^2 -     8 \sqrt{2} a_2 g_1^{R} m_{\eta^\prime}^2 M_{R}^2 m_K^2 - 
    8 \sqrt{2} a_4 g_1^{R} m_{\eta^\prime}^2 M_{R}^2 m_K^2  \nn\\ & +     8 a_2 (g_{8a}^{R} - g_{8b}^{R}) (m_{\eta^\prime}^2 - M_{R}^2)^2 m_K^2 - 
    4 a_4 (4 g_{8a}^{R} - g_{8b}^{R}) (m_{\eta^\prime}^2 - M_{R}^2)^2 m_K^2 +     4 \sqrt{2} a_2 g_1^{R} M_{R}^2 m_K^2 (M_{R}^2 - s)  \nn\\ & + 
    4 \sqrt{2} a_4 g_1^{R} M_{R}^2 m_K^2 (M_{R}^2 - s) -     \sqrt{2} a_2 g_1^{R} m_{\eta^\prime}^4 s - \sqrt{2} a_4 g_1^{R} m_{\eta^\prime}^4 s - 
    2 a_2 g_{8a}^{R} m_{\eta^\prime}^4 s + 4 a_4 g_{8a}^{R} m_{\eta^\prime}^4 s + 2 a_2 g_{8b}^{R} m_{\eta^\prime}^4 s  \nn\\ & -     a_4 g_{8b}^{R} m_{\eta^\prime}^4 s + 2 \sqrt{2} a_2 g_1^{R} m_{\eta^\prime}^2 M_{R}^2 s + 
    2 \sqrt{2} a_4 g_1^{R} m_{\eta^\prime}^2 M_{R}^2 s +     4 a_2 g_{8a}^{R} m_{\eta^\prime}^2 M_{R}^2 s - 8 a_4 g_{8a}^{R} m_{\eta^\prime}^2 M_{R}^2 s  \nn\\ & - 
    4 a_2 g_{8b}^{R} m_{\eta^\prime}^2 M_{R}^2 s + 2 a_4 g_{8b}^{R} m_{\eta^\prime}^2 M_{R}^2 s -     2 a_2 g_{8a}^{R} M_{R}^4 s + 4 a_4 g_{8a}^{R} M_{R}^4 s + 
    2 a_2 g_{8b}^{R} M_{R}^4 s - a_4 g_{8b}^{R} M_{R}^4 s \nn\\ & -     16 a_2 g_{8a}^{R} m_{\eta^\prime}^2 m_K^2 s + 32 a_4 g_{8a}^{R} m_{\eta^\prime}^2 m_K^2 s + 
    16 a_2 g_{8b}^{R} m_{\eta^\prime}^2 m_K^2 s - 8 a_4 g_{8b}^{R} m_{\eta^\prime}^2 m_K^2 s -     16 a_2 g_{8a}^{R} M_{R}^2 m_K^2 s  \nn\\ &   + 32 a_4 g_{8a}^{R} M_{R}^2 m_K^2 s + 
    16 a_2 g_{8b}^{R} M_{R}^2 m_K^2 s - 8 a_4 g_{8b}^{R} M_{R}^2 m_K^2 s -     \sqrt{2} a_2 g_1^{R} M_{R}^2 (M_{R}^2 - s) s  \nn\\ &  - 
    \sqrt{2} a_4 g_1^{R} M_{R}^2 (M_{R}^2 - s) s -     4 \sqrt{2} a_2 g_1^{R} m_K^2 (M_{R}^2 - s) s - 
    4 \sqrt{2} a_4 g_1^{R} m_K^2 (M_{R}^2 - s) s +     4 a_2 (g_{8a}^{R} - g_{8b}^{R}) m_{\eta^\prime}^2 s^2 \nn\\ &  + 4 a_2 (g_{8a}^{R} - g_{8b}^{R}) M_{R}^2 s^2 + 
    8 a_2 (g_{8a}^{R} - g_{8b}^{R}) m_K^2 s^2 -     2 a_4 (4 g_{8a}^{R} - g_{8b}^{R}) (m_{\eta^\prime}^2 + M_{R}^2 + 2 m_K^2) s^2 \nn\\ & + 
    \sqrt{2} a_2 g_1^{R} (M_{R}^2 - s) s^2  +     \sqrt{2} a_4 g_1^{R} (M_{R}^2 - s) s^2 + 4 a_4 g_{8a}^{R} s^3 - 
    2 a_2 (g_{8a}^{R} - g_{8b}^{R}) s^3 - a_4 g_{8b}^{R} s^3 \nn\\ & +     2 \sqrt{2} a_2 g_1^{R} m_{\eta^\prime}^2 s (-4 m_K^2 + s) +     2 \sqrt{2} a_4 g_1^{R} m_{\eta^\prime}^2 s (-4 m_K^2 + s)\big\}\,,
\end{align}

\begin{align}\label{Ampetappipi0}
M_0^{R, \chi,\eta^\prime \pi\pi}(s)=& -\frac{1}{3 \sqrt{3} F^3 s}
   \big\{2 \sqrt{2} a_2 g_1^{R} m_{\eta^\prime}^4 m_{\pi}^2 -      4 \sqrt{2} a_2 g_1^{R} m_{\eta^\prime}^2 M_{R}^2 m_{\pi}^2 + 
     2 a_2 (2 g_{8a}^{R} + g_{8b}^{R}) (m_{\eta^\prime}^2 - M_{R}^2)^2 m_{\pi}^2  \nn\\ & +      2 \sqrt{2} a_2 g_1^{R} M_{R}^2 m_{\pi}^2 (M_{R}^2 - s) + 
     \sqrt{2} a_2 g_1^{R} m_{\eta^\prime}^4 s - 2 \sqrt{2} a_2 g_1^{R} m_{\eta^\prime}^2 M_{R}^2 s -      a_2 (-2 g_{8a}^{R} - g_{8b}^{R}) (m_{\eta^\prime}^2 - M_{R}^2)^2 s  \nn\\ & + 
     8 a_2 (2 g_{8a}^{R} + g_{8b}^{R}) m_{\eta^\prime}^2 m_{\pi}^2 s -      16 a_2 (2 g_{8a}^{R} + g_{8b}^{R}) M_{R}^2 m_{\pi}^2 s +      36 a_2 (\sqrt{2} h_1^{R} - 2 h_8^{R}) m_K^2 m_{\pi}^2 s + 
     6 a_2 (11 \sqrt{2} h_1^{R} + 20 h_8^{R}) m_{\pi}^4 s  \nn\\ & -      24 a_4 (\sqrt{2} h_1^{R} + h_8^{R}) m_{\pi}^2 (-2 m_K^2 + m_{\pi}^2) s +      \sqrt{2} a_2 g_1^{R} M_{R}^2 (M_{R}^2 - s) s - 
     14 \sqrt{2} a_2 g_1^{R} m_{\pi}^2 (M_{R}^2 - s) s  \nn\\ &  -      2 a_2 (2 g_{8a}^{R} + g_{8b}^{R}) m_{\eta^\prime}^2 s^2 +      4 a_2 (2 g_{8a}^{R} + g_{8b}^{R}) M_{R}^2 s^2 + 
     14 a_2 (2 g_{8a}^{R} + g_{8b}^{R}) m_{\pi}^2 s^2 +      5 \sqrt{2} a_2 g_1^{R} (M_{R}^2 - s) s^2  \nn\\ & + 5 a_2 (-2 g_{8a}^{R} - g_{8b}^{R}) s^3 -      2 \sqrt{2} a_2 g_1^{R} m_{\eta^\prime}^2 s (-4 m_{\pi}^2 + s)\big\} \,,
\end{align}

\begin{align}\label{Ampetappipi2}
M_2^{R, \chi,\eta^\prime \pi\pi}(s)=& -\frac{1}{3 \sqrt{3} F^3 s}
   \big\{4 \sqrt{2} a_2 g_1^{R} m_{\eta^\prime}^4 m_{\pi}^2 -     8 \sqrt{2} a_2 g_1^{R} m_{\eta^\prime}^2 M_{R}^2 m_{\pi}^2 +     4 a_2 (2 g_{8a}^{R} + g_{8b}^{R}) (m_{\eta^\prime}^2 - M_{R}^2)^2 m_{\pi}^2 \nn\\ & + 
    4 \sqrt{2} a_2 g_1^{R} M_{R}^2 m_{\pi}^2 (M_{R}^2 - s) -     \sqrt{2} a_2 g_1^{R} m_{\eta^\prime}^4 s + 2 \sqrt{2} a_2 g_1^{R} m_{\eta^\prime}^2 M_{R}^2 s + 
    a_2 (-2 g_{8a}^{R} - g_{8b}^{R}) (m_{\eta^\prime}^2 - M_{R}^2)^2 s \nn\\ & -     8 a_2 (2 g_{8a}^{R} + g_{8b}^{R}) m_{\eta^\prime}^2 m_{\pi}^2 s -     8 a_2 (2 g_{8a}^{R} + g_{8b}^{R}) M_{R}^2 m_{\pi}^2 s - 
    \sqrt{2} a_2 g_1^{R} M_{R}^2 (M_{R}^2 - s) s -     4 \sqrt{2} a_2 g_1^{R} m_{\pi}^2 (M_{R}^2 - s) s \nn\\ & +     2 a_2 (2 g_{8a}^{R} + g_{8b}^{R}) m_{\eta^\prime}^2 s^2 + 
    2 a_2 (2 g_{8a}^{R} + g_{8b}^{R}) M_{R}^2 s^2 + 4 a_2 (2 g_{8a}^{R} + g_{8b}^{R}) m_{\pi}^2 s^2 +     \sqrt{2} a_2 g_1^{R} (M_{R}^2 - s) s^2 \nn\\ & + a_2 (-2 g_{8a}^{R} - g_{8b}^{R}) s^3 + 
    2 \sqrt{2} a_2 g_1^{R} m_{\eta^\prime}^2 s (-4 m_{\pi}^2 + s)\big\} \,,
\end{align}

\begin{align}\label{AmpetaKK0}
M_0^{R, \chi,\eta K\bar{K}}(s)=&  -\frac{1}{6 \sqrt{3} F F_K^2 s}
   \big\{2 \sqrt{2} a_1 g_1^{R} m_{\eta}^4 m_K^2 + 2 \sqrt{2} a_3 g_1^{R} m_{\eta}^4 m_K^2 -      4 \sqrt{2} a_1 g_1^{R} m_{\eta}^2 M_{R}^2 m_K^2 - 
     4 \sqrt{2} a_3 g_1^{R} m_{\eta}^2 M_{R}^2 m_K^2  \nn\\ & +  4 a_1 (g_{8a}^{R} - g_{8b}^{R}) (m_{\eta}^2 - M_{R}^2)^2 m_K^2 -      2 a_3 (4 g_{8a}^{R} - g_{8b}^{R}) (m_{\eta}^2 - M_{R}^2)^2 m_K^2 +      2 \sqrt{2} a_1 g_1^{R} M_{R}^2 m_K^2 (M_{R}^2 - s) \nn\\ & +    2 \sqrt{2} a_3 g_1^{R} M_{R}^2 m_K^2 (M_{R}^2 - s) +      \sqrt{2} a_1 g_1^{R} m_{\eta}^4 s + \sqrt{2} a_3 g_1^{R} m_{\eta}^4 s +      2 a_1 g_{8a}^{R} m_{\eta}^4 s - 4 a_3 g_{8a}^{R} m_{\eta}^4 s - 2 a_1 g_{8b}^{R} m_{\eta}^4 s  \nn\\ & + 
     a_3 g_{8b}^{R} m_{\eta}^4 s - 2 \sqrt{2} a_1 g_1^{R} m_{\eta}^2 M_{R}^2 s -      2 \sqrt{2} a_3 g_1^{R} m_{\eta}^2 M_{R}^2 s - 
     4 a_1 g_{8a}^{R} m_{\eta}^2 M_{R}^2 s + 8 a_3 g_{8a}^{R} m_{\eta}^2 M_{R}^2 s  \nn\\ & + 
     4 a_1 g_{8b}^{R} m_{\eta}^2 M_{R}^2 s - 2 a_3 g_{8b}^{R} m_{\eta}^2 M_{R}^2 s +      2 a_1 g_{8a}^{R} M_{R}^4 s - 4 a_3 g_{8a}^{R} M_{R}^4 s - 
     2 a_1 g_{8b}^{R} M_{R}^4 s + a_3 g_{8b}^{R} M_{R}^4 s  \nn\\ & - 
          20 a_1 g_{8a}^{R} m_{\eta}^2 m_K^2 s + 4 a_3 g_{8a}^{R} m_{\eta}^2 m_K^2 s +      20 a_1 g_{8b}^{R} m_{\eta}^2 m_K^2 s - 28 a_3 g_{8b}^{R} m_{\eta}^2 m_K^2 s + 
     4 a_1 g_{8a}^{R} M_{R}^2 m_K^2 s  \nn\\ & + 28 a_3 g_{8a}^{R} M_{R}^2 m_K^2 s -      4 a_1 g_{8b}^{R} M_{R}^2 m_K^2 s + 20 a_3 g_{8b}^{R} M_{R}^2 m_K^2 s + 
     6 \sqrt{2} a_3 h_1^{R} (24 m_K^4 - 6 m_K^2 m_{\pi}^2 - m_{\pi}^4) s  \nn\\ &  -      24 a_1 h_8^{R} (2 m_K^4 + m_K^2 m_{\pi}^2 - m_{\pi}^4) s -      24 a_3 h_8^{R} (6 m_K^4 - 6 m_K^2 m_{\pi}^2 + m_{\pi}^4) s + 
     6 \sqrt{2} a_1 h_1^{R} (4 m_K^4 + 20 m_K^2 m_{\pi}^2 + m_{\pi}^4) s \nn\\ & +      \sqrt{2} a_1 g_1^{R} M_{R}^2 (M_{R}^2 - s) s +      \sqrt{2} a_3 g_1^{R} M_{R}^2 (M_{R}^2 - s) s - 
     14 \sqrt{2} a_1 g_1^{R} m_K^2 (M_{R}^2 - s) s  \nn\\ & -      14 \sqrt{2} a_3 g_1^{R} m_K^2 (M_{R}^2 - s) s -      10 a_3 g_{8a}^{R} m_{\eta}^2 s^2 + 14 a_1 (g_{8a}^{R} - g_{8b}^{R}) m_{\eta}^2 s^2 + 
     16 a_3 g_{8b}^{R} m_{\eta}^2 s^2 + 2 a_3 g_{8a}^{R} M_{R}^2 s^2 \nn\\ &  -      10 a_1 (g_{8a}^{R} - g_{8b}^{R}) M_{R}^2 s^2 - 14 a_3 g_{8b}^{R} M_{R}^2 s^2 -      20 a_3 g_{8a}^{R} m_K^2 s^2 - 8 a_1 (g_{8a}^{R} - g_{8b}^{R}) m_K^2 s^2 -      22 a_3 g_{8b}^{R} m_K^2 s^2  \nn\\ & + 5 \sqrt{2} a_1 g_1^{R} (M_{R}^2 - s) s^2 +      5 \sqrt{2} a_3 g_1^{R} (M_{R}^2 - s) s^2 + 2 a_3 g_{8a}^{R} s^3 + 
     8 a_1 (g_{8a}^{R} - g_{8b}^{R}) s^3 + 13 a_3 g_{8b}^{R} s^3  \nn\\ & -      2 \sqrt{2} a_1 g_1^{R} m_{\eta}^2 s (-4 m_K^2 + s) -      2 \sqrt{2} a_3 g_1^{R} m_{\eta}^2 s (-4 m_K^2 + s)\big\}\,,
\end{align}

\begin{align}\label{AmpetaKK2}
M_2^{R, \chi,\eta K\bar{K}}(s)=& -\frac{1}{6 \sqrt{3} F F_K^2 s}
   \big\{4 \sqrt{2} a_1 g_1^{R} m_{\eta}^4 m_K^2 +     4 \sqrt{2} a_3 g_1^{R} m_{\eta}^4 m_K^2 -     8 \sqrt{2} a_1 g_1^{R} m_{\eta}^2 M_{R}^2 m_K^2 - 
    8 \sqrt{2} a_3 g_1^{R} m_{\eta}^2 M_{R}^2 m_K^2  \nn\\ & +     8 a_1 (g_{8a}^{R} - g_{8b}^{R}) (m_{\eta}^2 - M_{R}^2)^2 m_K^2 - 
    4 a_3 (4 g_{8a}^{R} - g_{8b}^{R}) (m_{\eta}^2 - M_{R}^2)^2 m_K^2 +     4 \sqrt{2} a_1 g_1^{R} M_{R}^2 m_K^2 (M_{R}^2 - s)  \nn\\ & + 
    4 \sqrt{2} a_3 g_1^{R} M_{R}^2 m_K^2 (M_{R}^2 - s) -     \sqrt{2} a_1 g_1^{R} m_{\eta}^4 s - \sqrt{2} a_3 g_1^{R} m_{\eta}^4 s - 
    2 a_1 g_{8a}^{R} m_{\eta}^4 s + 4 a_3 g_{8a}^{R} m_{\eta}^4 s + 2 a_1 g_{8b}^{R} m_{\eta}^4 s  \nn\\ & -     a_3 g_{8b}^{R} m_{\eta}^4 s + 2 \sqrt{2} a_1 g_1^{R} m_{\eta}^2 M_{R}^2 s + 
    2 \sqrt{2} a_3 g_1^{R} m_{\eta}^2 M_{R}^2 s +     4 a_1 g_{8a}^{R} m_{\eta}^2 M_{R}^2 s - 8 a_3 g_{8a}^{R} m_{\eta}^2 M_{R}^2 s  \nn\\ & - 
    4 a_1 g_{8b}^{R} m_{\eta}^2 M_{R}^2 s + 2 a_3 g_{8b}^{R} m_{\eta}^2 M_{R}^2 s -     2 a_1 g_{8a}^{R} M_{R}^4 s + 4 a_3 g_{8a}^{R} M_{R}^4 s + 
    2 a_1 g_{8b}^{R} M_{R}^4 s - a_3 g_{8b}^{R} M_{R}^4 s \nn\\ & -     16 a_1 g_{8a}^{R} m_{\eta}^2 m_K^2 s + 32 a_3 g_{8a}^{R} m_{\eta}^2 m_K^2 s + 
    16 a_1 g_{8b}^{R} m_{\eta}^2 m_K^2 s - 8 a_3 g_{8b}^{R} m_{\eta}^2 m_K^2 s -     16 a_1 g_{8a}^{R} M_{R}^2 m_K^2 s  \nn\\ &   + 32 a_3 g_{8a}^{R} M_{R}^2 m_K^2 s + 
    16 a_1 g_{8b}^{R} M_{R}^2 m_K^2 s - 8 a_3 g_{8b}^{R} M_{R}^2 m_K^2 s -     \sqrt{2} a_1 g_1^{R} M_{R}^2 (M_{R}^2 - s) s  \nn\\ &  - 
    \sqrt{2} a_3 g_1^{R} M_{R}^2 (M_{R}^2 - s) s -     4 \sqrt{2} a_1 g_1^{R} m_K^2 (M_{R}^2 - s) s - 
    4 \sqrt{2} a_3 g_1^{R} m_K^2 (M_{R}^2 - s) s +     4 a_1 (g_{8a}^{R} - g_{8b}^{R}) m_{\eta}^2 s^2 \nn\\ &  + 4 a_1 (g_{8a}^{R} - g_{8b}^{R}) M_{R}^2 s^2 + 
    8 a_1 (g_{8a}^{R} - g_{8b}^{R}) m_K^2 s^2 -     2 a_3 (4 g_{8a}^{R} - g_{8b}^{R}) (m_{\eta}^2 + M_{R}^2 + 2 m_K^2) s^2 \nn\\ & + 
    \sqrt{2} a_1 g_1^{R} (M_{R}^2 - s) s^2  +     \sqrt{2} a_3 g_1^{R} (M_{R}^2 - s) s^2 + 4 a_3 g_{8a}^{R} s^3 - 
    2 a_1 (g_{8a}^{R} - g_{8b}^{R}) s^3 - a_3 g_{8b}^{R} s^3 \nn\\ & +     2 \sqrt{2} a_1 g_1^{R} m_{\eta}^2 s (-4 m_K^2 + s) +     2 \sqrt{2} a_3 g_1^{R} m_{\eta}^2 s (-4 m_K^2 + s)\big\}\,,
\end{align}

\begin{align}\label{Ampetapipi0}
M_0^{R,\chi, \eta \pi\pi}(s)=& -\frac{1}{3 \sqrt{3} F^3 s}
   \big\{2 \sqrt{2} a_1 g_1^{R} m_{\eta}^4 m_{\pi}^2 -      4 \sqrt{2} a_1 g_1^{R} m_{\eta}^2 M_{R}^2 m_{\pi}^2 + 
     2 a_1 (2 g_{8a}^{R} + g_{8b}^{R}) (m_{\eta}^2 - M_{R}^2)^2 m_{\pi}^2  \nn\\ & +      2 \sqrt{2} a_1 g_1^{R} M_{R}^2 m_{\pi}^2 (M_{R}^2 - s) + 
     \sqrt{2} a_1 g_1^{R} m_{\eta}^4 s - 2 \sqrt{2} a_1 g_1^{R} m_{\eta}^2 M_{R}^2 s -      a_1 (-2 g_{8a}^{R} - g_{8b}^{R}) (m_{\eta}^2 - M_{R}^2)^2 s  \nn\\ & + 
     8 a_1 (2 g_{8a}^{R} + g_{8b}^{R}) m_{\eta}^2 m_{\pi}^2 s -      16 a_1 (2 g_{8a}^{R} + g_{8b}^{R}) M_{R}^2 m_{\pi}^2 s +      36 a_1 (\sqrt{2} h_1^{R} - 2 h_8^{R}) m_K^2 m_{\pi}^2 s + 
     6 a_1 (11 \sqrt{2} h_1^{R} + 20 h_8^{R}) m_{\pi}^4 s  \nn\\ & -      24 a_3 (\sqrt{2} h_1^{R} + h_8^{R}) m_{\pi}^2 (-2 m_K^2 + m_{\pi}^2) s +      \sqrt{2} a_1 g_1^{R} M_{R}^2 (M_{R}^2 - s) s - 
     14 \sqrt{2} a_1 g_1^{R} m_{\pi}^2 (M_{R}^2 - s) s  \nn\\ &  -      2 a_1 (2 g_{8a}^{R} + g_{8b}^{R}) m_{\eta}^2 s^2 +      4 a_1 (2 g_{8a}^{R} + g_{8b}^{R}) M_{R}^2 s^2 + 
     14 a_1 (2 g_{8a}^{R} + g_{8b}^{R}) m_{\pi}^2 s^2 +      5 \sqrt{2} a_1 g_1^{R} (M_{R}^2 - s) s^2  \nn\\ & + 5 a_1 (-2 g_{8a}^{R} - g_{8b}^{R}) s^3 -      2 \sqrt{2} a_1 g_1^{R} m_{\eta}^2 s (-4 m_{\pi}^2 + s)\big\} \,,
\end{align}

\begin{align}\label{Ampetapipi2}
M_2^{R,\chi, \eta \pi\pi}(s)=& -\frac{1}{3 \sqrt{3} F^3 s}
   \big\{4 \sqrt{2} a_1 g_1^{R} m_{\eta}^4 m_{\pi}^2 -     8 \sqrt{2} a_1 g_1^{R} m_{\eta}^2 M_{R}^2 m_{\pi}^2 +     4 a_1 (2 g_{8a}^{R} + g_{8b}^{R}) (m_{\eta}^2 - M_{R}^2)^2 m_{\pi}^2 \nn\\ & + 
    4 \sqrt{2} a_1 g_1^{R} M_{R}^2 m_{\pi}^2 (M_{R}^2 - s) -     \sqrt{2} a_1 g_1^{R} m_{\eta}^4 s + 2 \sqrt{2} a_1 g_1^{R} m_{\eta}^2 M_{R}^2 s + 
    a_1 (-2 g_{8a}^{R} - g_{8b}^{R}) (m_{\eta}^2 - M_{R}^2)^2 s \nn\\ & -     8 a_1 (2 g_{8a}^{R} + g_{8b}^{R}) m_{\eta}^2 m_{\pi}^2 s -     8 a_1 (2 g_{8a}^{R} + g_{8b}^{R}) M_{R}^2 m_{\pi}^2 s - 
    \sqrt{2} a_1 g_1^{R} M_{R}^2 (M_{R}^2 - s) s -     4 \sqrt{2} a_1 g_1^{R} m_{\pi}^2 (M_{R}^2 - s) s \nn\\ & +     2 a_1 (2 g_{8a}^{R} + g_{8b}^{R}) m_{\eta}^2 s^2 + 
    2 a_1 (2 g_{8a}^{R} + g_{8b}^{R}) M_{R}^2 s^2 + 4 a_1 (2 g_{8a}^{R} + g_{8b}^{R}) m_{\pi}^2 s^2 +     \sqrt{2} a_1 g_1^{R} (M_{R}^2 - s) s^2 \nn\\ & + a_1 (-2 g_{8a}^{R} - g_{8b}^{R}) s^3 + 
    2 \sqrt{2} a_1 g_1^{R} m_{\eta}^2 s (-4 m_{\pi}^2 + s)\big\} \,,
\end{align}
where the kaon decay constant, $F_K = 0.113$ GeV, has been employed.
For the sake of brevity, we express the $S$- and $D$-waves of amplitudes of $ X(2370)/X(1835)/\eta_c \to \eta[\eta^\prime] \pi\pi(K\bar{K}) $ processes in an unified form above, in which ``$R$'' can denote different initial state $X(2370)/X(1835)/\eta_c$. Note that since $\eta_c$ is a flavor-singlet state, the couplings satisfy $ g_{8a}^{\eta_c}=g_{8b}^{\eta_c}=h_{8}^{\eta_c}=0$.

\end{appendix}

\end{onecolumngrid}

\bibliography{refs}

\end{document}